\documentclass[pra,showpacs,twocolumn,amsmath,amssymb,floatfix]{revtex4}

\usepackage{graphicx}

\newcommand{\pdagger}{{\phantom{\dagger}}}
\newcommand{\dt}{\Delta\tau}
\newcommand{\reff}[1]{Fig.\ \ref{fig:#1}}
\newcommand{\reffl}[1]{Figure\ \ref{fig:#1}}

\newcommand{\myparagraph}[1]{{\it #1} -- }

\begin{document}

	\title{Mott transitions in ternary flavor mixtures of ultracold fermions on optical lattices}

\author{E.~V.~Gorelik}
%\email{gorelike@uni-mainz.de}
\affiliation{Institute of Physics, Johannes Gutenberg University, 55099 Mainz, Germany}
\author{N.~Bl\"umer}
\affiliation{Institute of Physics, Johannes Gutenberg University, 55099 Mainz, Germany}

\date{\today}

  \begin{abstract}
	  Ternary flavor mixtures of ultracold fermionic atoms in an optical lattice are studied
    in the case of equal, repulsive on-site interactions $U>0$. The corresponding SU(3) invariant 
    Hubbard model is solved numerically exactly within dynamical mean-field theory using
    multigrid Hirsch-Fye quantum Monte Carlo simulations. We establish Mott transitions close 
    to integer filling at low temperatures and show that the associated signatures in
    the compressibility and pair occupancy persist to high temperatures, i.e., should be
    accessible to experiments. In addition, we present spectral functions and discuss the
    properties of a novel ``semi-compressible'' state observed for large $U$ near half filling.
  \end{abstract}
  \maketitle

%%%%%%%%%%%%%%%%%%%%%%%%%%%%%%%%%%%%%%%%%%%%%%%%%%%%%%%%%%%%%%%%%%%%%%%%
% Introduction
%%%%%%%%%%%%%%%%%%%%%%%%%%%%%%%%%%%%%%%%%%%%%%%%%%%%%%%%%%%%%%%%%%%%%%%%

  Starting with the achievement of Bose Einstein condensation in 1995, the creation and
  study of quantum degenerate atomic gases has led to discoveries with enormous impact
  far beyond atomic physics \cite{Bloch08_RMP}. In particular, atomic gases 
can be driven to the strongly correlated regime by switching on optical lattices and/or 
   using Feshbach resonances. In the
  bosonic case, the localized Mott phase is accessible already in 
  single-component (lattice) systems \cite{Greiner02}. With the recent observation
  \cite{Esslinger08_Nature,Bloch08_ferm} of Mott transitions in balanced 2-flavor (i.e.
  2 hyperfine state) mixtures of fermionic $^{40}K$ atoms with repulsive interactions,
  such systems are now established as highly tunable {\it quantum simulators} of
  condensed matter \cite{Bloch08_Science}.

  At the same time, ultracold atoms offer new degrees of freedom: 
  In contrast to the electronic case, with only two spin
  states, fermionic atoms have large hyperfine multiplets. Early experiments
  \cite{Regal03} involving three hyperfine states of $^{40}K$ have prompted theoretical
  investigations of 3-flavor mixtures on optical lattices. These studies have focussed
  on the case of pair-wise equal {\it attractive} interactions which may induce 
  trionic phases and exotic
  superfluidity with a 3-component order parameter, somewhat analogous to
  QCD \cite{Honerkamp_Rapp}. 
  Within the last year, balanced 3-flavor $^6$Li mixtures have
  been trapped and studied across Feshbach resonances \cite{Ottenstein_PRL08,Huckans08}.
  Soon, such systems, both with repulsive and attractive interactions, should be
  realized also in optical lattices.

%%%%%%%%%%%%%%%%%%%%%%%%%%%%%%%%%%%%%%%%%%%%%%%%%%%%%%%%%%%%%%%%%%%%%%%%%%% 

  In this paper, we explore the properties of fermionic 3-flavor mixtures on optical
  lattices in the case of {\it repulsive} interactions. While the many ordering
  patterns conceivable in such systems will certainly warrant extensive studies at some
  point, ordering phenomena have escaped experimental detection even in the simpler
  2-flavor case so far. We will, thus, concentrate on the %(nonperturbative) 
  Mott physics of {\it homogeneous phases}. 
  Note that this regime is particularly challenging since
  it requires nonperturbative theoretical approaches. 
 
%%%%%%%%%%%%%%%%%%%%%%%%%%%%%%%%%%%%%%%%%%%%%%%%%%%%%%%%%%%%%%%%%%%%%%%%
% Method
%%%%%%%%%%%%%%%%%%%%%%%%%%%%%%%%%%%%%%%%%%%%%%%%%%%%%%%%%%%%%%%%%%%%%%%%

  Ternary mixtures of fermions on an optical 
  lattice can be modeled via the Hubbard-type Hamiltonian
  \begin{equation}\label{Hubb_mod}
    \hat{H} =\! \sum_{\langle ij\rangle ,\alpha}  t_\alpha\, \hat{c}^{\dag}_{i\alpha}
  \hat{c}^\pdagger_{j\alpha} 
  + \!\!\sum_{i, \alpha<\alpha'} U_{\alpha\alpha'}\, \hat{n}_{i\alpha} \hat{n}_{i\alpha'}
  \,- \sum_{i, \alpha}\mu_\alpha\, \hat{n}_{i\alpha}.
  \end{equation}
  Here, $\langle ij\rangle$ denotes nearest-neighbor sites; $\alpha\in\{1,2,3\}$ labels
  the fermionic flavors, $t$ parameterizes the hopping, $U$ the on-site Coulomb
  interaction, and $\mu$ is the chemical potential.
  In Eq.\ (\ref{Hubb_mod}), we have neglected higher Bloch bands and the confining
  potential. 
  For fermions of a single species (in the vibrational ground state),
  $t_\alpha\approx t_{\alpha'}$. Following the literature, we will also assume pairwise
  equal interactions and study the SU(3) symmetric limit $t_\alpha\equiv t,\,
  U_{\alpha\alpha'}\equiv U,\, \mu_\alpha\equiv \mu+U$; by this definition, 
  particle-hole symmetry corresponds to a sign change in $\mu$.

  This system is treated within dynamical mean-field theory (DMFT) which retains the
  dynamics of local correlations \cite{Kotliar_Vollhardt}. %Throughout this work, we will use the 
The semi-elliptic ``Bethe''
  density of states with bandwidth $W=4t^*$ mimics \cite{fn:lattice} a simple
  cubic lattice (with $t=t^*/\sqrt{6}$); as usual in DMFT studies, the scale is set by the scaled hopping amplitude $t^*=\sqrt{Z}t$ for coordination number $Z$. %(with $t=1/\sqrt{6}$). 
The DMFT impurity problem is solved using the Hirsch-Fye quantum
  Monte Carlo algorithm (HF-QMC) \cite{Hirsch86}, first in the conventional form, with
  imaginary-time discretization $\dt\le 0.8/U$; the validity of these results is later
  verified and unbiased results are obtained using the numerically exact multigrid
  implementation \cite{Bluemer_multigrid}. 
  Note that already conventional HF-QMC is competitive with continuous-time QMC
  \cite{Bluemer_efficiency}; the multigrid variant is even more efficient.
%  Both methods are competitive with continous-time QMC \cite{Bluemer_efficiency}.  
  All QMC based methods share the advantage,
  compared to numerical renormalization group (NRG) approaches, of being reliable
  at the experimentally relevant (elevated) temperatures.
%  is not directly accessible in QMC simulations.

%entropies are
%  not directly accessible in QMC simulations.
  
%%%%%%%%%%%%%%%%%%%%%%%%%%%%%%%%%%%%%%%%%%%%%%%%%%%%%%%%%%%%%%%%%%%%%%%% 
% Numerical  results %%%%%%%%%%%%%%%%%%%%%%%%%%%%%%%%%%%%%%%%%%%%%%%%%%%%%%%%%%%%%%%%%%%%%%%%

 \myparagraph{Results at low temperature} 
  Theoretical studies of electronic 1-band models are often restricted to half filling
%  , i.e., obtained at fixed 
  ($\mu=0$). Mott metal-insulator transition are, then, signaled
  by kinks or jumps in thermodynamic properties (as a function of interaction $U$ and/or
  temperature $T$) or, more directly, by the opening of a gap in the local spectral
  function $A(\omega)$. In the present 3-flavor case, genuine Mott physics can be
  expected only far away from particle-hole symmetry; consequently, the chemical
  potential $\mu$ is an essential additional parameter. In the following, we will choose
  a relatively low temperature $T=t^*/20$ and explore the ($U$, $\mu$) space with
  particular emphasis on properties that are most accessible in quantum gas experiments
  and which are related to Mott physics. Temperature effects and spectral functions are
  to be discussed later.

%\newpage
%%%%%%%%%%%%%%%%%%%%%%%%%%%%%%%%%%%%%%%%%%%%%%%%%%%%%%%%%%%%%%%%%%%%%%%%%%%%%%%%%
\begin{figure} 
	\includegraphics[width=\columnwidth]{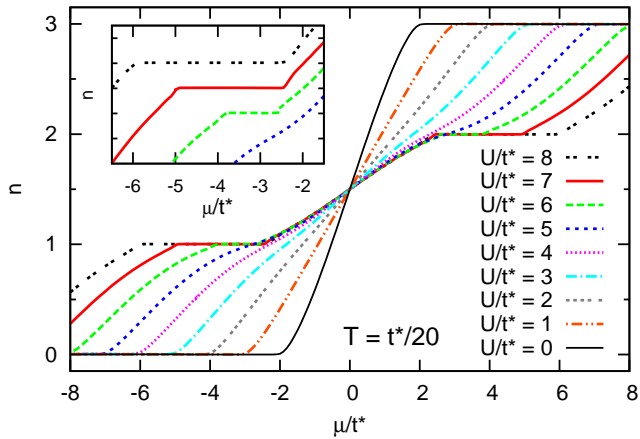}
\caption{(Color online) HF-QMC estimates of particle density $n(\mu)$ at $T=t^*/20$ for various on-site interactions $U$.
Plateaus at integer
filling indicate localized Mott phases (for $U/t^*\ge 6$). In the inset, curves
are shifted (in steps of 0.1) for clarity.}\label{fig:n_vs_mu} 
\end{figure}
%%%%%%%%%%%%%%%%%%%%%%%%%%%%%%%%%%%%%%%%%%%%%%%%%%%%%%%%%%%%%%%%%%%%%%%%%%%%%%%%%
  \reffl{n_vs_mu} shows the filling $n=\sum_{\alpha} \langle \hat{n}_\alpha\rangle$
  versus the chemical potential $\mu$ for a range of on-site
	interactions $U$. Initially, for $U=0$, $n$ varies smoothly and rapidly with $\mu$
	from an empty band ($n=0$) at $\mu/t^*\le -2$ to a full band ($n=3$) at $\mu/t^*\ge 2$. With
	increasing $U$, the slope generally decreases, but the curves remain smooth until, for
	$U/t^*\ge 6$, plateaus develop at integer filling $n=1$, $n=2$ which signal the
	onset of localized Mott phases (and correspond to gaps in the spectral function).
	The particle-hole symmetry around $\mu=0$ is evident; from now on, we will only   
	consider
	$\mu\le 0$. The plateaus appear perfectly flat even at the magnified scale of the
	inset of \reff{n_vs_mu} (for $n=1$); tendencies towards hysteresis are seen
	for $U/t^*=6$ (at $\mu/t^*\approx-2.5$). We will argue
	below that $T=t^*/20$ is very close to the critical temperature $T^*$ of a second-order
	Mott critical end point (cf.\ \reff{FSMT_multi_B}).

  From our results, detailed predictions for experimental measurements 
  can be derived within the local density approximation (LDA). For a specific trap
  potential and atom number, the data of \reff{n_vs_mu}, then, translates
  to occupancies $n_i$ of the inhomogeneous system,  integrated
	 column densities and cloud sizes. Analogous approaches have been used in the Bloch-Rosch
	 collaboration for establishing the Mott phase in the 2-flavor case
	 \cite{Helmes_PRL08,Bloch08_ferm}.
	
%%%%%%%%%%%%%%%%%%%%%%%%%%%%%%%%%%%%%%%%%%%%%%%%%%%%%%%%%%%%%%%%%%%%%%%%%%%%%%%%%
  \begin{figure} 
	\includegraphics[width=\columnwidth]{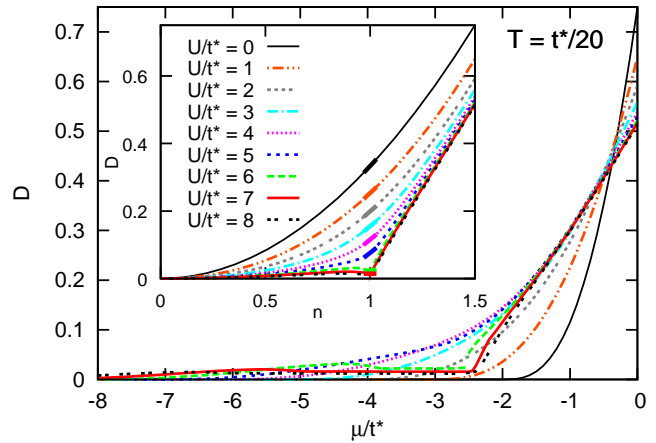}
  \caption{(Color online) HF-QMC estimates of pair occupancy $D$ at $T=t^*/20$ vs.\
  chemical potential $\mu$ (main panel) and vs.\ density $n$ (inset), respectively.
  The suppression of $D$ with increasing $U$ is clearly enhanced on the Mott
  plateau at $n\approx 1$. }\label{fig:D_vs_mu_vs_n} \end{figure}
%%%%%%%%%%%%%%%%%%%%%%%%%%%%%%%%%%%%%%%%%%%%%%%%%%%%%%%%%%%%%%%%%%%%%%%%%%%%%%%%%

	 A fundamental property of Mott phases is the suppression of density
   fluctuations; in the 2-flavor case these are usually quantified by the
   double occupancy $\langle \hat{n}_{i\uparrow} \hat{n}_{i\downarrow}\rangle$. In the
   present 3-flavor context, we consider the more general {\it pair occupancy}
   $D=\sum_{\alpha<\alpha'} \langle \hat{n}_{i\alpha} \hat{n}_{i\alpha'}\rangle$,
   which retains the energy relation $E=E_{\text{kin}} + U D$ \cite{fn:D}.
	 As seen in the main panel of \reff{D_vs_mu_vs_n}, $D$ depends strongly  on $\mu$
   and $U$; the dependence on $\mu$ is mostly monotonic, except for the vicinity of
   plateaus (for $U/t^*\ge 6$). The impact of $U$ is best understood at fixed
   density $n$ (see inset of \reff{D_vs_mu_vs_n}): starting from the noninteracting
   limit $U=0$, where $D=n^2/3$, $D$ is suppressed with increasing $U$ at all $n$.
   This suppression is strongly enhanced at $n=1$ for $U/t^*\ge 6$.
   Fortunately, the probabilities for n-fold occupancy can be easily measured
   experimentally. In fact, this technique has been used by the Esslinger group for
   detecting Mott phases in the 2-flavor case via changes in the slopes $dD/dn$ close to
   integer filling \cite{Esslinger08_Nature}; analogous slopes are marked by thick line
   segments in the inset of \reff{D_vs_mu_vs_n}.

	One interesting feature of \reff{n_vs_mu} not discussed so far is that the slopes
	$dn/d\mu$ seem to saturate in the range $-2\lesssim \mu/t^* \lesssim 2$ for large $U$. The
	corresponding compressibility $\kappa = dn/d\mu$ \cite{Kotliar_PRL02}, computed via
	numerical differentiation of the data of \reff{n_vs_mu}, is shown in \reff{dn_dmu}. 
	Note that -- in contrast to expressions such as $n^{-1} dn/d\mu$ -- this definition of $\kappa$
	retains the particle-hole symmetry ($\kappa(-\mu)=\kappa(\mu)$) and fulfils a sum rule
	($\int_{-\infty}^\infty \kappa(\mu) d\mu=3$), just as the spectral function. Indeed,
	$\kappa(\mu)$ reduces to the slightly broadened semi-elliptic spectral function for
	$U=0$ (thick solid line in \reff{dn_dmu}). 
	With increasing $U$, the height of this
		central peak is rapidly reduced and the weight of $\kappa(\mu)$ is shifted towards larger $|\mu|$,
		with a subpeak at $|\mu|\approx U+t^*$ developing for $U/t^*\gtrsim 2$. The
		intermediate minima are initially smooth; their positions coincide with integer filling
		as is apparent from the curves $\kappa(n)$ shown in the inset of \reff{dn_dmu}. Finally,
		for $U/t^*\gtrsim 6$, sharp gaps appear; these Mott plateaus [with $\kappa(\mu)\approx 0$]
		collapse to points in the representation $\kappa(n)$. 
		
		Coming back to the saturation of slopes $dn/d\mu$, let us first note that the peak value
		$\kappa(\mu=0)$ reaches a finite limit of about $0.2/t^*$ for large $U$ (and low $T$), i.e.,
		scales with the inverse of the (only remaining) energy scale $W\propto t^*$.
		Even more importantly, the variation of $\kappa(\mu)$ within the central compressible phase
		decreases with increasing $U$ until $\kappa$ is nearly constant for $U/t^*\ge 8$ [in 
		contrast with the edge phase ($n<1$) where $\kappa t^*$ varies between 0 and about 0.3]. This
		additional plateau marks a new ``semi-compressible'' state  sandwiched between 2 
		equivalent Mott phases which is specific to SU($2M+1$)
		systems ($M\ge 1$) \cite{FlorensGeorges_04} and should be easy to probe 
		experimentally.
		
	 %%%%%%%%%%%%%%%%%%%%%%%%%%%%%%%%%%%%%%%%%%%%%%%%%%%%%%%%%%%%%%%%%%%%%%%%%%%%%%%%%
		 \begin{figure} 
			\includegraphics[width=\columnwidth]{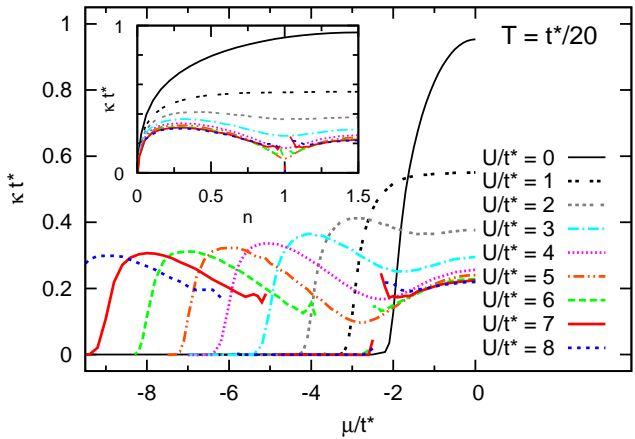}
	  \caption{(Color online) HF-QMC estimates of compressibility $\kappa=dn/d\mu$
	  at $T=t^*/20$ as a function of chemical potential $\mu$ (main panel) and density $n$
	  (inset), respectively. 
	  }\label{fig:dn_dmu} \end{figure}
		%%%%%%%%%%%%%%%%%%%%%%%%%%%%%%%%%%%%%%%%%%%%%%%%%%%%%%%%%%%%%%%%%%%%%%%%%%%%%%%%%	

	%%%%%%%%%%%%%%%%%%%
	\myparagraph{Impact of temperature}
	While experiments with fermionic atomic gases in optical lattices offer
	much greater flexibility in varying interaction and hopping parameters
	(compared to solid state experiments), they are currently restricted by initial temperatures not much lower than the Fermi temperature $T_{\text{F}}$. On
	the other hand, true Mott transitions and/or ordering phenomena typically occur on
	energy scales that are 1-2 orders of magnitude below the kinetic energy scales (such as
	$T_{\text{F}}$ or the bandwidth $W$). Therefore, it is essential to study the systems of
	interest also at elevated temperatures and, if possible, to identify signatures of
	Mott transitions which persist to temperatures much higher than the
	critical temperatures, i.e., signal well-defined crossover regions. At the same time,
	good estimates of the critical temperature(s) $T^*$ of the Mott transitions are 
	desirable. 
	
 	%%%%%%%%%%%%%%%%%%%%%%%%%%%%%%%%%%%%%%%%%%%%%%%%%%%%%%%%%%%%%%%%%%%%%%%%%%%%%%%%%
 \begin{figure} 
	\includegraphics[width=\columnwidth]{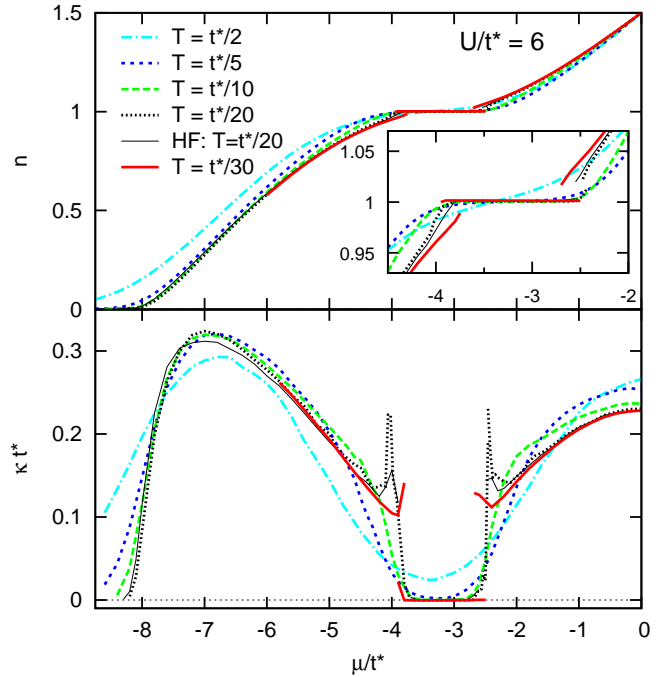}
  \caption{(Color online) Multigrid HF-QMC estimates of density $n(\mu)$ (top panel) and
  compressibility $\kappa(\mu)=dn(\mu)/d\mu$ (bottom panel) at interaction $U/t^*=6$  for various temperatures $T$. The strong
  suppression of $\kappa$ at $n\approx 1$ survives up to $T\approx t^*/5 \gg T^*\approx
  t^*/20$. For comparison, thin solid lines indicate HF-QMC data for $T=t^*/20$ (cf.\ Figs.\
  \ref{fig:n_vs_mu} and \ref{fig:dn_dmu}). }\label{fig:FSMT_multi_B} \end{figure}
	%%%%%%%%%%%%%%%%%%%%%%%%%%%%%%%%%%%%%%%%%%%%%%%%%%%%%%%%%%%%%%%%%%%%%%%%%%%%%%%%%

In \reff{FSMT_multi_B}, we present multigrid HF-QMC estimates (thick lines) of $n(\mu)$
(top panel) and $\kappa(\mu)$ (lower panel) for a wide range of temperatures at a
relatively strong interaction $U/t^*=6$.
Clearly, the temperature effects are small across wide regions of $\mu$, with deviations
in $n$ typically of the order $10^{-2}$ for $T\le t^*/5$. However, extended hysteresis regions at the plateau edges are observed only at the lowest temperature, $T=t^*/30$. 
At $T=t^*/20$,
coexistence is hardly significant (around $\mu/t^*\approx -2.5$) while
 the compressibility is strongly enhanced in both transition
regions, as expected very close to $2^{\text{nd}}$ order critical end points \cite{Kotliar_PRL02}. Consequently, we estimate the critical temperature as 
$T^*=(0.05\pm 0.01)t^*$. 
At higher temperatures, all curves are smooth. Nevertheless, a Mott
plateau can be identified up to $T=t^*/5$: in the central region $\mu/t^*\approx -3.3$, the
compressibility is still reduced by two orders of magnitudes compared to off-plateau
regions. This is good news
for experimentalists: a factor of 4 above the critical temperature [roughly corresponding to an entropy $s=\log(3)$ per particle],
the (remnant of the) Mott plateau should still be well visible. Even at $T=t^*/2$, a full order
of magnitude above $T^*$, the compressibility is stronger reduced, in the Mott crossover
region, than at the subcritical interaction $U/t^*=5$ for low $T$ (cf.\ \reff{dn_dmu}).
A similar picture emerges for the pair occupancy
 (not shown): both the plateaus and the nonmonotonic behavior in $D(\mu)$
   persist up to $T\lesssim t^*/5$.
 
Also shown in \reff{FSMT_multi_B} are conventional HF-QMC estimates for $T=t^*/20$ (and $\dt t^*=0.12$; thin
solid lines). The good agreement with the numerically exact multigrid data (thick solid
lines) shows that the discretization effects of the former method are hardly significant; 
thus, the data shown in \reff{n_vs_mu} -- \ref{fig:dn_dmu} are reliable.

\myparagraph{Spectra}
Finally, let us discuss an experimentally more challenging observable: the local spectral function $A(\omega)$.
%%%%%%%%%%%%%%%%%%%%%%%%%%%%%%%%%%%%%%%%%%%%%%%%%%%%%%%%%%%%%%%%%%%%%%%%%%%%%%%%%
\begin{figure} 
	\includegraphics[width=\columnwidth]{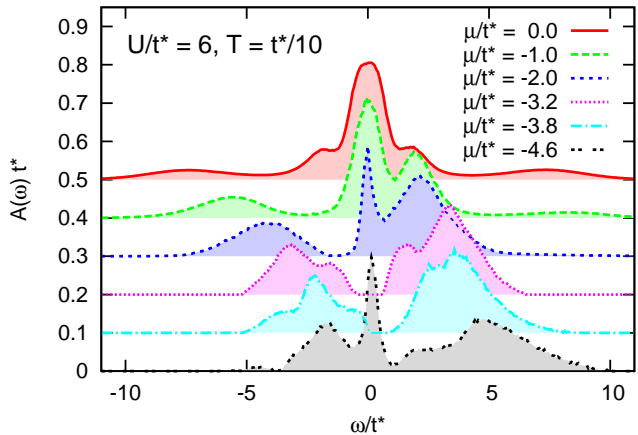}
\caption{(Color online) Multigrid HF-QMC estimates of spectral function $A(\omega)$ for $T=t^*/10$, $U/t^*=6$ at different
values of $\mu$, corresponding to densities $n=1.5, 1.27, 1.07,
1.00, 0.9996$, and $0.91$ (from top to bottom). 
% With decreasing $n$, the upper Hubbard band disappears; the lower Hubbard band 
% acquires a weight of 1/3 at the Mott transition.
}\label{fig:spectra} \end{figure}
%%%%%%%%%%%%%%%%%%%%%%%%%%%%%%%%%%%%%%%%%%%%%%%%%%%%%%%%%%%%%%%%%%%%%%%%%%%%%%%%%
\reffl{spectra} shows estimates obtained via maximum entropy analytic continuation
 from numerically exact multigrid HF-QMC Green functions for $U/t^*=6$. 
At half filling (top curve), $A(\omega)$ is symmetric and shows a
rough 3-peak structure (with peaks at $\omega=0$ and $\omega/t^*\approx \pm 7$), reminiscent of the
familiar 1-band-2-spin correlated metallic phase. However, important differences arise
in the 3-flavor case: (i) the Hubbard bands, although well separated, carry little
weight (of about $10\%$ each); (ii) the central peak (at $|\omega/t^*|\lesssim 4$) has a three-fold substructure; (iii) the quasi-particle peak remains for arbitrary $U$ (with only $10\%$ reduction in weight for $5\le U/t^*\le 10$; not shown). 
%Very roughly, the total
%structure might be understood as a 4-fold atomic multiplet plus a central quasi-particle
%resonance; in the correlated regime, quantum fluctuations of the system (with
%average filling $n=1.5$) will mostly involve occupations 1 or 2. Fluctuations
%towards occupations 0 and 3, which contribute to the Hubbard bands in $A(\omega)$, are,
%therefore, suppressed. 

With decreasing filling, triply occupied sites become even less important: the upper
Hubbard band vanishes for $\mu/t^*\lesssim -2$. In contrast, the lower band rapidly acquires
weight: about $26\%$ for $\mu/t^*=-1$, while the quasiparticle peak shrinks at nearly constant peak height ($A(0) t^*\approx 0.3$). It disappears when $\mu$
reaches the Mott plateau: at $\mu/t^*=-3.2$, a well-defined central gap has opened. The
remaining peaks have centers of masses at $\omega=-2.8 t^*$ and $\omega =3.2 t^*$, respectively; the
difference exactly equals the interaction $U$, just as one would expect in the
half-filled 2-spin case. However, in the present 3-flavor case (also at $n=1$), the
weights are strongly asymmetric: $1/3$ and $2/3$ for the lower and upper Hubbard band
(of which the latter has formed from the central peak at $n=1.5$), respectively. For
further decreased $\mu$, a quasiparticle peak reappears from the lower subband. 

\myparagraph{Conclusion} 
In this first nonperturbative study of the repulsive SU(3)
symmetric Hubbard model, we have thoroughly explored its homogeneous phases at low and
intermediate temperatures in the $(U,\mu)$ space. We have shown that signatures of Mott 
transitions close to integer filling persist to elevated temperatures, facilitating
experimental observation of strong correlation phenomena, e.g., in balanced 3-flavor 
mixtures of $^{40}$K (with roughly equal positive scattering lengths at small magnetic 
field \cite{Regal03}) or $^6$Li \cite{Ottenstein_PRL08,Huckans08} on optical lattices; 
systems with Raman excited states \cite{MuellerBloch_PRL2007} provide an interesting alternative.

We have found peculiar features in spectra and a novel ``semi-compressible'' phase with asymptotically universal $\kappa$ (independent of $U$, $T$, and $\mu$ or $n$), which is clearly linked with the presence of adjacent {\it equivalent} Mott lobes. Experimental studies of this new physics might be challenging, but are within reach of existing techniques.

%worthwhile, but  tasks for experimentalists.

%All of these predictions can be tested with existing experimental methods (including RF spectroscopy).

%Apart from verifying ``old'' Mott physics in the novel 3-flavor systems (with
%connections to QCD), experimentalists should focus on the new physics, i.e., study
%the ``semi-compressible'' phase ($1<n<2$),  and access dynamic information, e.g.,
%using RF spectroscopy.

Our work could be extended towards full quantitative correspondence with upcoming
experiments by explicit inclusion of the trap potential, within LDA (similar to recent studies
\cite{Bloch08_ferm,De_Leo_PRL08,Scarola_PRL09} for 2-flavor systems) or beyond \cite{Helmes_PRL08,Snoek_NJP08}; also, ordered phases \cite{Honerkamp_PRL04,Snoek_NJP08} could be considered. It seems, however, more urgent
to investigate which properties of SU(3) symmetric systems survive
for unequal interactions (with less severe 3-body losses \cite{Ottenstein_PRL08,Huckans08}),
when flavor-selective Mott physics \cite{Knecht05} may be expected.

  We thank I.\ Bloch, P.G.J.\ van Dongen, W.\ Hofstetter, \'{A}.\ Rapp, and U.\ Schneider for discussions.
  Support by the DFG within the TR 49 and by the John von Neumann Institute for Computing is
  gratefully acknowledged.


\begin{thebibliography}{16}
\bibitem{fn:D} The probability of double (but not triple) occupancy is $\tilde{D}\equiv D-3\langle n_{i1} n_{i2} n_{i3}\rangle$; however, for $\mu\le 0$, $D\approx \tilde{D}$.
\end{thebibliography}
\end{document}